\documentclass[11pt]{article}

\usepackage[margin=1in]{geometry}
\usepackage{amsmath,amssymb,amsfonts}
\usepackage{booktabs}
\usepackage{graphicx}
\usepackage{natbib}
\usepackage{hyperref}
\usepackage{microtype}

\hypersetup{
  colorlinks=true,
  linkcolor=blue,
  citecolor=blue,
  urlcolor=blue
}

\newcommand{\PWR}{\mathrm{PWR}}
\newcommand{\WR}{\mathrm{WR}}
\newcommand{\Prb}{\Pr}
\newcommand{\E}{\mathbb{E}}
\DeclareMathOperator{\logit}{logit}

\title{Probabilistic Win Ratio Method for Hierarchical Composite Endpoints with Coarsened Outcomes}

\author{
Lei Li\thanks{LunarAI LLC. Email: \texttt{lei.li@lunarai.llc}. Corresponding author.}
\and
Jing Lei\thanks{Incyte Corporation. Email: \texttt{JLei@incyte.com}.}
\and
Yuexiao Dong\thanks{Fox School of Business, Temple University. Email: \texttt{yuexiao.dong@temple.edu}.}
}

\date{Preprint}

\begin{document}
\maketitle

\begin{abstract}
The win ratio is increasingly used to analyze prioritized composite endpoints in clinical trials, but standard implementations rely on deterministic pairwise comparisons and can perform poorly in the presence of censoring and endpoint-specific missingness. In such settings, unresolved comparisons are often treated as ties, leading to loss of efficiency and potentially biased inference, particularly when lower-priority outcomes are incompletely observed. We propose the probabilistic win ratio (PWR), a framework for estimating the classical win ratio under coarsened observation. The PWR replaces deterministic pairwise decisions with conditional probabilities of win, loss, or tie given the observed data, allowing partially observed comparisons to contribute fractionally while being explicitly penalized according to their uncertainty. Comparisons with greater coarsening receive smaller effective weight, whereas fully observed comparisons contribute as in the classical analysis, preserving the clinical priority structure. When outcomes are fully observed, the PWR reduces exactly to the standard win ratio estimator. Simulation studies show that the PWR maintains low bias and mean squared error across a range of censoring and missingness scenarios. Two clinical trial case studies illustrate complementary data regimes, demonstrating calibration in near-complete data and stability under substantial right censoring.
\end{abstract}

\noindent\textbf{Keywords:} probabilistic win ratio; hierarchical composite endpoints; coarsened data; censoring; missingness.

\section{Introduction}

Composite endpoints are widely used in clinical trials to improve statistical efficiency by capturing multiple clinically relevant outcomes within a single analysis. In many disease areas, however, component endpoints differ substantially in clinical importance, with outcomes such as death or irreversible organ failure considered more critical than nonfatal events. This has motivated the use of prioritized composite endpoints, in which patient outcomes are compared hierarchically according to a prespecified order of clinical importance.

The win ratio proposed by \citet{pocock2012win} and its generalizations \citep{buyse2022gpc, peron2018extension, gasparyan2021adjusted} have become prominent tools for analyzing such prioritized endpoints. These methods compare patients pairwise across treatment arms, resolving each comparison at the highest-priority layer where a difference is observed. The win ratio has been successfully applied in cardiovascular, oncology, and critical care trials and is increasingly recommended when conventional time-to-first-event analyses fail to reflect clinical priorities. Despite their interpretability and appeal, existing win-ratio-based methods face important practical challenges when outcomes are coarsened by censoring and layer-specific missingness.

In many modern trials, higher-priority outcomes such as mortality are nearly fully ascertained, whereas lower-priority outcomes, often nonfatal, recurrent, or resource-intensive to collect, are more susceptible to incomplete follow-up and missingness. Classical win ratio analyses typically treat partially observed comparisons as ties, discarding available information and potentially reducing efficiency. Extensions based on inverse-probability-of-censoring weighting (IPCW) and proportional win fraction regression have been proposed to address censoring \citep{dong2020ipcw, mao2021proportional}, but these approaches rely on correct specification of the censoring process and do not directly exploit partial information available at unresolved layers. Multiple-imputation strategies are flexible but introduce additional modeling assumptions and computational complexity, and may be sensitive to model misspecification.

In this paper, we propose a probabilistic win ratio framework for estimating the classical win ratio under coarsened observation. Rather than relying on deterministic pairwise decisions, the PWR replaces unresolved comparisons with probability-based contributions derived from subject-specific outcome models. For each hierarchical layer, marginal models are fitted and converted into conditional distributions that respect each subject's observed coarsening state. Pairwise win, loss, and tie probabilities are then computed, typically by numerical integration, and uncertainty is propagated to lower-priority layers solely through tie probabilities from higher-priority layers. We derive an explicit layer-wise decomposition that clarifies how individual components contribute to the overall comparison. Importantly, the PWR targets the same population estimand as the classical win ratio and reduces exactly to the deterministic win ratio when outcomes are fully observed.

Through simulation studies, we evaluate the finite-sample performance of the PWR relative to the classical win ratio, IPCW-based win ratio, and multiple-imputation benchmarks under varying degrees of censoring and layer-specific missingness. We further illustrate the method using two clinical trial datasets chosen to represent complementary data regimes: a near-complete dataset that serves as a calibration example and a heavily censored dataset that demonstrates the stability of the PWR under substantial coarsening.

The remainder of the paper is organized as follows. Section~\ref{sec:method} introduces the PWR and develops a general framework for hierarchical comparisons under censoring and layer-specific missingness, with a specialization to prioritized survival endpoints using conditional survival modeling and tie-probability propagation. Section~\ref{sec:simulation} presents a two-layer simulation study comparing the PWR with classical, IPCW-based, and multiple-imputation win ratio methods across a range of censoring and missingness scenarios. Section~\ref{sec:cases} reports two case studies based on the primary biliary cirrhosis trial and the HF-ACTION trial. Section~\ref{sec:discussion} concludes with discussion and future directions.

\section{Method}
\label{sec:method}

\subsection{General framework under coarsened outcomes}

Consider a two-arm randomized study comparing a treatment group ($A=1$) and a control group ($A=0$). For subject $i$, let
\[
Y_i = \{Y_i^{(1)},\ldots,Y_i^{(K)}\}
\]
denote a vector of latent outcome components ordered by prespecified clinical priority, with lower indices indicating higher priority. The components may be of mixed type, such as continuous, binary, or time-to-event, and no common scale or universal direction of desirability is assumed across components. Let $X_i$ denote baseline covariates measured prior to treatment assignment. We assume that $X_i$ is fully observed throughout this paper.

\subsubsection{Win ratio hierarchical comparison rule}

Clinical preference between two subjects is determined by a fixed hierarchical comparison rule. For each layer $\ell \in \{1,\ldots,K\}$, let $\succ_\ell$ denote a clinically specified strict preference relation on the support of $Y^{(\ell)}$, and let $\sim_\ell$ denote the corresponding indifference relation. Given a treated-control pair $(i,j)$, the rule compares $Y_i^{(1)}$ and $Y_j^{(1)}$ under $(\succ_1,\sim_1)$; if $Y_i^{(1)} \sim_1 Y_j^{(1)}$, it proceeds to layer 2, and so on, until either a strict preference is determined or all layers are tied.

This induces a latent pairwise comparison indicator
\[
W_{ij}\in\{-1,0,1\},
\]
where
\[
W_{ij} =
\begin{cases}
1, & \text{if } Y_i^{(\ell)} \succ_\ell Y_j^{(\ell)} \text{ at the first decisive layer } \ell,\\
-1, & \text{if } Y_j^{(\ell)} \succ_\ell Y_i^{(\ell)} \text{ at the first decisive layer } \ell,\\
0, & \text{if } Y_i^{(\ell)} \sim_\ell Y_j^{(\ell)} \text{ for all } \ell=1,\ldots,K.
\end{cases}
\]

In principle, the hierarchical comparison rule induces a clear clinical preference between any two subjects through the latent outcome vectors $Y_i$ and $Y_j$. In practice, however, these outcomes are rarely fully observed. Time-to-event components may be right-censored, lower-priority outcomes may be missing, and the first decisive layer cannot always be deterministically identified from the observed data.

The classical win ratio applies the hierarchical rule directly to observed outcomes, treating unresolved comparisons as ties and proceeding mechanically to lower-priority layers. As a consequence, uncertainty arising from censoring or missingness is collapsed into deterministic decisions, which can distort the effective weighting of outcome components and reduce efficiency.

The PWR addresses this gap by recognizing that under coarsened observation, the comparison outcome $W_{ij}$ is itself unobserved. Rather than forcing a deterministic classification, the PWR estimates the win ratio by evaluating conditional win, loss, and tie probabilities given the observed baseline covariates and previous layer information, and by propagating uncertainty through the hierarchy.

\subsubsection{The classical win ratio}

Let $(i,j)$ denote a randomly selected treated-control pair, where $i$ is drawn uniformly from the treated group and $j$ is drawn uniformly from the control group, independently of each other. Under complete observation of $Y_i$ and $Y_j$, define the population win and loss probabilities $\Prb(W_{ij}=1)$ and $\Prb(W_{ij}=-1)$. The population win ratio is
\begin{equation}
\theta = \frac{\Prb(W_{ij}=1)}{\Prb(W_{ij}=-1)},
\label{eq:theta}
\end{equation}
with the convention that $\theta=\infty$ if $\Prb(W_{ij}=-1)=0$.

Under complete observation, the win and loss probabilities can be estimated empirically by counting pairwise wins and losses across all treated-control pairs. Let $n_1$ and $n_0$ denote the numbers of treated and control subjects, respectively. The classical win ratio estimator is
\begin{equation}
\widehat{\theta}_{\WR} =
\frac{
 n_1^{-1} n_0^{-1}\sum_{i:A_i=1}\sum_{j:A_j=0} I(W_{ij}=1)
}{
 n_1^{-1} n_0^{-1}\sum_{i:A_i=1}\sum_{j:A_j=0} I(W_{ij}=-1)
}.
\label{eq:classic}
\end{equation}

\subsubsection{Coarsened outcomes and a probabilistic representation}

For each subject $i$ and layer $k=0,1,\ldots,K$, define the nested sigma-fields
\[
\mathcal{F}_{i,0}=\sigma(X_i),\qquad
\mathcal{F}_{i,1:k}
=\sigma\{X_i,\text{ coarsened information on }Y_i^{(1)},\ldots,Y_i^{(k)}\},
\]
and set $\mathcal{F}_i=\mathcal{F}_{i,1:K}$. For a treated-control pair $(i,j)$, define
\[
\mathcal{F}_{ij,1:k}=\sigma(\mathcal{F}_{i,1:k},\mathcal{F}_{j,1:k}),
\qquad
\mathcal{F}_{ij}=\mathcal{F}_{ij,1:K}.
\]
Intuitively, $\mathcal{F}_{ij,1:k}$ contains all observed information from layers $1,\ldots,k$ that may affect how the pairwise comparison is interpreted up to layer $k$ and whether the rule proceeds to lower layers.

Under coarsening, $W_{ij}$ is generally not observable from $\mathcal{F}_{ij}$, but its conditional distribution given $\mathcal{F}_{ij}$ is well defined:
\[
\Prb(W_{ij}=w\mid \mathcal{F}_{ij}),\qquad w\in\{-1,0,1\}.
\]
By the law of iterated expectations, the win ratio in \eqref{eq:theta} admits the identity
\begin{equation}
\theta =
\frac{\E\{\Prb(W_{ij}=1\mid \mathcal{F}_{ij})\}}
{\E\{\Prb(W_{ij}=-1\mid \mathcal{F}_{ij})\}}.
\label{eq:prob_representation}
\end{equation}
Equation~\eqref{eq:prob_representation} motivates estimating $\theta$ under coarsening by replacing the unobservable conditional probabilities with model-based estimates constructed from the observed data.

\subsection{Estimation under coarsened outcomes}

The PWR estimator $\widehat{\theta}_{\PWR}$ is a plug-in estimator of the classical win ratio estimand that replaces the unknown conditional probabilities in \eqref{eq:prob_representation} by model-based estimates constructed from the observed coarsened data.

\subsubsection{Layer-wise objects and tie propagation}

For $k=1,\ldots,K$, let $R_{ij,k}$ denote the latent event indicator that the hierarchical rule reaches layer $k$, meaning that layers $1,\ldots,k-1$ are tied under the latent hierarchical ordering. Conditional on reaching layer $k$, define the layer-$k$ conditional comparison probabilities
\[
p^+_{ij,k}=\Prb(\text{win}_k\mid \mathcal{F}_{ij,1:k},R_{ij,k}=1),\quad
p^-_{ij,k}=\Prb(\text{loss}_k\mid \mathcal{F}_{ij,1:k},R_{ij,k}=1),
\]
and
\[
p^0_{ij,k}=\Prb(\text{tie}_k\mid \mathcal{F}_{ij,1:k},R_{ij,k}=1),
\]
with $p^+_{ij,k}+p^-_{ij,k}+p^0_{ij,k}=1$ almost surely on $\{R_{ij,k}=1\}$.

Because the endpoint hierarchy is prespecified and clinically ordered, higher-priority outcomes are defined, recorded, and interpreted independently of lower-priority outcomes. Accordingly, we assume that information recorded at lower-priority layers does not provide additional information about higher-priority comparisons beyond what is already contained in the observed history up to that layer. Formally, for each $k$,
\begin{align}
\Prb(R_{ij,k}=1\mid \mathcal{F}_{ij})
&=\Prb(R_{ij,k}=1\mid \mathcal{F}_{ij,1:k-1}),\nonumber\\
\Prb(\text{outcome}^u_k\mid \mathcal{F}_{ij},R_{ij,k}=1)
&=\Prb(\text{outcome}^u_k\mid \mathcal{F}_{ij,1:k},R_{ij,k}=1),
\label{eq:no_feedback}
\end{align}
for $u\in\{+,-,0\}$.

Define reach weights by
\[
q_{ij,1}=1,\qquad
q_{ij,k}=\Prb(R_{ij,k}=1\mid \mathcal{F}_{ij,1:k-1}),\quad k\geq 2.
\]
Under the hierarchical rule, reaching layer $k+1$ requires reaching layer $k$ and tying at layer $k$, so
\[
q_{ij,k+1}=q_{ij,k}p^0_{ij,k},\qquad k=1,\ldots,K-1,
\]
or equivalently $q_{ij,k}=\prod_{\ell=1}^{k-1}p^0_{ij,\ell}$ for $k\geq 2$. With these definitions and assumption \eqref{eq:no_feedback}, the overall conditional win and loss probabilities admit the decomposition
\[
\Prb(W_{ij}=1\mid \mathcal{F}_{ij})=\sum_{k=1}^K q_{ij,k}p^+_{ij,k},
\qquad
\Prb(W_{ij}=-1\mid \mathcal{F}_{ij})=\sum_{k=1}^K q_{ij,k}p^-_{ij,k}.
\]

\subsubsection{PWR estimator}

Let $\widehat p^+_{ij,k}$, $\widehat p^-_{ij,k}$, and $\widehat p^0_{ij,k}$ denote estimators of the layer-wise conditional comparison probabilities, obtained from fitted working models at each layer that are compatible with the observed coarsening mechanism. Define estimated reach weights by
\[
\widehat q_{ij,1}=1,\qquad
\widehat q_{ij,k+1}=\widehat q_{ij,k}\widehat p^0_{ij,k},
\]
and estimated pairwise conditional win and loss probabilities by
\[
\widehat\pi^+_{ij}=\sum_{k=1}^K \widehat q_{ij,k}\widehat p^+_{ij,k},
\qquad
\widehat\pi^-_{ij}=\sum_{k=1}^K \widehat q_{ij,k}\widehat p^-_{ij,k}.
\]
The PWR estimator is
\begin{equation}
\widehat{\theta}_{\PWR} =
\frac{
 n_1^{-1} n_0^{-1}\sum_{i:A_i=1}\sum_{j:A_j=0}\widehat\pi^+_{ij}
}{
 n_1^{-1} n_0^{-1}\sum_{i:A_i=1}\sum_{j:A_j=0}\widehat\pi^-_{ij}
}.
\label{eq:pwr}
\end{equation}
Under full observation, the fitted conditional probabilities degenerate to deterministic decisions and $\widehat\pi^+_{ij}=I(W_{ij}=1)$ and $\widehat\pi^-_{ij}=I(W_{ij}=-1)$, so \eqref{eq:pwr} reduces exactly to \eqref{eq:classic}. Under coarsened observation, the estimated conditional probabilities take values in $(0,1)$ as determined by the working models, so partially observed comparisons contribute fractionally rather than as full wins, losses, or ties.

\subsection{Specialization to survival outcomes}

The general framework above applies to arbitrary coarsened outcome vectors. We now specialize to hierarchical survival endpoints, where each component is a latent event time.

We consider a hierarchy of $K$ time-to-event endpoints ordered by decreasing clinical priority. For subject $i$, let
\[
T_i=\{T_i^{(1)},\ldots,T_i^{(K)}\}
\]
denote the corresponding latent event times. The highest-priority layer need not correspond specifically to death; rather, it may represent any fatal, terminal, or absorbing endpoint such that, once it occurs, subsequent lower-priority outcomes are no longer meaningfully observable or clinically relevant.

\subsubsection{Survival-layer probabilities under coarsening}

Conditional on $\mathcal{F}_{i,1:k}$, define the conditional survival functions
\[
S_{i,k}(t\mid \mathcal{F}_{i,1:k})
=\Prb(T_i^{(k)}>t\mid \mathcal{F}_{i,1:k}),\qquad
S_{j,k}(t\mid \mathcal{F}_{j,1:k})
=\Prb(T_j^{(k)}>t\mid \mathcal{F}_{j,1:k}),
\]
with corresponding conditional densities when required. By construction, these condition on both the observed layer-$k$ state and all higher-priority information.

Given that the hierarchical comparison reaches layer $k$, define at horizon $\tau$ the conditional layer-$k$ win, loss, and tie probabilities as
\begin{align}
p^-_{ij,k}(\tau)
&=\Prb(T_i^{(k)}<T_j^{(k)},T_i^{(k)}\leq \tau
  \mid \mathcal{F}_{ij,1:k},R_{ij,k}=1),\\
p^+_{ij,k}(\tau)
&=\Prb(T_j^{(k)}<T_i^{(k)},T_j^{(k)}\leq \tau
  \mid \mathcal{F}_{ij,1:k},R_{ij,k}=1),\\
p^0_{ij,k}(\tau)
&=\Prb(T_i^{(k)}>\tau,T_j^{(k)}>\tau
  \mid \mathcal{F}_{ij,1:k},R_{ij,k}=1).
\end{align}
Under continuous-time survival models, ties arise not from equality of event times but from joint survival beyond the layer-specific horizon $\tau$.

Assuming conditional independence across subjects given their own observed histories, the layer-$k$ probabilities admit the representation
\begin{align}
p^-_{ij,k}(\tau)
&=\int_0^\tau f_{i,k}(t\mid \mathcal{F}_{i,1:k})
  S_{j,k}(t\mid \mathcal{F}_{j,1:k})\,dt,\\
p^+_{ij,k}(\tau)
&=\int_0^\tau f_{j,k}(t\mid \mathcal{F}_{j,1:k})
  S_{i,k}(t\mid \mathcal{F}_{i,1:k})\,dt,\\
p^0_{ij,k}(\tau)
&=S_{i,k}(\tau\mid \mathcal{F}_{i,1:k})
  S_{j,k}(\tau\mid \mathcal{F}_{j,1:k}).
\end{align}

In practice, the functions $f_{i,k}$ and $S_{i,k}$ are obtained from working models that condition on the observed history, including baseline covariates $X_i$ and coarsened information from higher-priority layers. These models may be parametric, semi-parametric, or otherwise flexible, and are used to construct the layer-wise conditional comparison probabilities.

Thus $\widehat{\theta}_{\PWR}$ is a covariate-conditional estimator of the classical win ratio in the sense that pairwise win and loss contributions are evaluated through conditional probabilities given the observed covariates and coarsening states. In addition, $\widehat{\theta}_{\PWR}$ averages these conditional probabilities over the empirical distribution of $(X_i,X_j)$ induced by random sampling of treated and control subjects. Modeling assumptions influence efficiency and finite-sample performance through the estimation of these conditional probabilities, but under appropriate assumptions on the coarsening mechanism they do not alter the underlying estimand.

Identification of the layer-wise conditional probabilities requires assumptions on the coarsening mechanism. Throughout, we assume that right censoring and missingness are conditionally independent of the latent event times given the observed history, including baseline covariates and higher-priority layer information. This corresponds to a missing-at-random or independent censoring assumption with respect to the filtration generated by $\mathcal{F}_{i,1:k}$. Under this assumption, the conditional survival functions and eligibility probabilities appearing in the working representation are identifiable from the observed data. A formal statement of the coarsening assumptions is provided in Appendix~\ref{app:regularity}.

\subsubsection{Layer-wise behavior under survival coarsening}

Table~\ref{tab:pwr_rules} summarizes the qualitative behavior of the PWR at a given layer. The table describes how the comparison at layer $k$ is handled depending on the observed states for the treatment and control subjects.

\begin{table}[htbp]
\centering
\caption{Summary of PWR rules at hierarchical layer $k$ under coarsening and layer-specific missingness.}
\label{tab:pwr_rules}
\begin{tabular}{p{0.25\linewidth}p{0.65\linewidth}}
\toprule
Treatment status & PWR rule\\
\midrule
Event & Deterministic if the control has an event or is alive at $\tau$; otherwise resolved probabilistically with $\Prb(\mathrm{win}_k)\in(0,1)$.\\
Alive at $\tau$ & Win if the control has an event; tie if the control is alive at $\tau$; otherwise mixed with positive probabilities of win and tie.\\
Censored & Resolved probabilistically if the control has an event; otherwise mixed with positive probabilities of win, loss, and tie and possible progression to layer $k+1$.\\
Missing & All outcomes are model-based, with uncertainty penalized through fractional contributions and possible progression to layer $k+1$.\\
\bottomrule
\end{tabular}
\end{table}

Several features are worth emphasizing. First, when both subjects experience the event and the ordering is observed, the probabilistic rule reduces to the classical deterministic decision. Second, censoring or missingness does not automatically force a tie; instead, it induces a non-degenerate conditional distribution over win, loss, and tie outcomes. Finally, uncertainty propagates to lower-priority layers only through the probability of a tie at the current layer, preserving the clinical priority structure while accounting for coarsened observation.

\subsection{Variance estimation and inference}

Inference for the PWR is based on the large-sample behavior of pairwise win and loss contributions. Since $\widehat{\theta}_{\PWR}$ is defined as a ratio of averaged pairwise conditional probabilities, its sampling variability is nontrivial and depends on the joint distribution of treated-control pairs as well as the working models used to estimate layer-specific probabilities.

In practice, we estimate variability using the nonparametric bootstrap at the subject level. Subjects are resampled with replacement within treatment groups, preserving the original randomization ratio. For each bootstrap sample, $\widehat{\theta}_{\PWR}$ is recomputed in its entirety, including refitting all working models and recalculating pairwise win, loss, and tie probabilities. Let $\widehat{\theta}_{\PWR}^{*b}$ denote the estimate from bootstrap replicate $b=1,\ldots,B$.

Inference is performed on the log win ratio scale. The bootstrap standard error is estimated as the empirical standard deviation of $\log(\widehat{\theta}_{\PWR}^{*b})$, and confidence intervals are obtained using percentile or normal-approximation intervals on the log scale and then exponentiated back to the win ratio scale.

\subsection{Asymptotic properties}

The estimator $\widehat{\theta}_{\PWR}$ can be expressed as a smooth functional of empirical averages of pairwise quantities, where each treated-control pair contributes a bounded conditional win or loss probability. As such, the estimator falls within the general class of two-sample U-statistics with estimated nuisance parameters, similar to those studied by \citet{lehmann1963robust} and \citet{bebu2016large}.

Under the regularity conditions stated in Appendix~\ref{app:regularity}, including independent sampling of subjects, a fixed and bounded analysis horizon, positivity of censoring survival functions, and conditional independence of censoring and missingness given the observed history, $\widehat{\theta}_{\PWR}$ is consistent for $\theta$. Consistency follows from uniform convergence of the estimated layer-specific survival functions and stability of the hierarchical tie-propagation mapping under coarsened observation.

For large samples, the estimator admits a standard first-order linearization as a ratio of U-statistics with plug-in nuisance estimators. Consequently, $\widehat{\theta}_{\PWR}$ exhibits root-$n$ convergence and asymptotically Gaussian behavior under mild smoothness conditions on the working survival models and the coarsening mechanism. The asymptotic variance depends on second-order moments of the pairwise conditional win and loss contributions and on the estimation error of the nuisance parameters.

\section{Simulation study}
\label{sec:simulation}

We consider a two-layer simulation study mimicking clinical trials. The first layer is death and the second layer is time to first hospitalization. Both endpoints are time-to-event endpoints. The simulation study was designed to reflect key features of prioritized composite endpoints commonly used in clinical trials, while isolating the inferential challenges that motivate the proposed PWR method.

In practice, higher-priority outcomes such as mortality are typically well observed, whereas lower-priority outcomes are more prone to missingness and incomplete follow-up. Right censoring due to administrative study end or dropout is also ubiquitous. Accordingly, event times across layers were generated jointly to allow dependence without enforcing deterministic ordering, censoring was introduced through a combination of administrative censoring and random dropout, and missingness was imposed selectively on lower-priority layers only among subjects who remained eligible for comparison at that layer.

\subsection{Data-generating mechanism}

For each subject $i$, we generated latent event times $(T_i^{(1)},T_i^{(2)})$ corresponding to death and time to first hospitalization. Treatment assignment $A_i\in\{0,1\}$ was generated independently with $\Prb(A_i=1)=0.5$. Baseline covariates $(X_{i1},X_{i2})$ were generated independently as standard normal random variables. An observed baseline risk group indicator $G_i\in\{0,1\}$ was generated independently with $\Prb(G_i=1)=0.5$.

To induce dependence across layers, we generated $(U_{i1},U_{i2})$ from a Gaussian copula with correlation $\rho$ and set $(U_{i1},U_{i2})=(\Phi(V_{i1}),\Phi(V_{i2}))$, where $(V_{i1},V_{i2})$ is bivariate normal with mean zero and correlation $\rho$. Latent event times were then generated from Weibull models by the inverse-CDF construction
\[
T_i^{(k)}
=
\left\{\frac{-\log(U_{ik})}{\lambda_{ik}\exp(\eta_{ik})}\right\}^{1/\kappa_k},
\qquad k\in\{1,2\},
\]
where $\kappa_1=1$ and $\kappa_2=0.6$. We used class-specific baseline intensities $\lambda_{ik}=\lambda_{k0}$ if $G_i=0$ and $\lambda_{ik}=\lambda_{k1}$ if $G_i=1$. The linear predictors were
\[
\eta_{i1}=\beta^\top X_i+0.4G_i,\qquad
\eta_{i2}=\beta^\top X_i+0.5G_i+A_i(-0.1-1.2G_i),
\]
so that treatment has no effect on death but reduces hospitalization risk with a stronger effect in the higher-risk subgroup. We set $\beta=(0.5,-0.3)$, $(\lambda_{10},\lambda_{11})=(0.02,0.04)$, and $(\lambda_{20},\lambda_{21})=(0.10,1.60)$.

Administrative censoring and dropout were represented by an independent censoring time $C_i$. Subjects were administratively followed to $t_{\max}$, and a proportion $p_D$ experienced dropout. For dropouts, $C_i$ was sampled uniformly on $[f_Dt_{\max},t_{\max}]$, while completers had $C_i=t_{\max}$. Observed death outcomes were
\[
\widetilde T_i^{(1)}=\min(T_i^{(1)},C_i),\qquad
\widetilde \Delta_i^{(1)}=I(T_i^{(1)}\leq C_i).
\]
Hospitalization was subject to the hierarchical stopping rule at death:
\[
\widetilde T_i^{(2)}=\min(T_i^{(2)},T_i^{(1)},C_i),\qquad
\widetilde \Delta_i^{(2)}=I\{T_i^{(2)}\leq \min(T_i^{(1)},C_i)\}.
\]

Missingness was imposed only on the hospitalization layer. Missingness was allowed only among subjects who were observed alive and uncensored through a landmark time $L$, i.e., among those with $\widetilde T_i^{(1)}\geq L$. For eligible subjects, we generated a missingness indicator $M_i\in\{0,1\}$ with
\[
\Prb(M_i=1\mid X_i,G_i,A_i,\widetilde T_i^{(1)}\geq L)
=
\logit^{-1}\{\alpha_0+0.3X_{i1}+1.2G_i+1.2(1-A_i)\},
\]
and calibrated $\alpha_0$ so that the marginal missingness rate among eligible subjects matched a target level $p_M$. When $M_i=1$, we set $(\widetilde T_i^{(2)},\widetilde\Delta_i^{(2)})$ to missing. No missingness was introduced in the oracle truth calculation.

All methods targeted the win ratio at a fixed analysis horizon $\tau$. For each simulated dataset, we constructed an oracle dataset by truncating latent event times at $\tau$ and enforcing hierarchy:
\[
T_{i,\tau}^{(1)}=\min(T_i^{(1)},\tau),\qquad
\Delta_{i,\tau}^{(1)}=I(T_i^{(1)}\leq \tau),
\]
\[
T_{i,\tau}^{(2)}=\min(T_i^{(2)},T_i^{(1)},\tau),\qquad
\Delta_{i,\tau}^{(2)}=I\{T_i^{(2)}\leq \min(T_i^{(1)},\tau)\}.
\]
The classical two-layer win ratio on this fully observed oracle dataset is denoted $\WR_{\mathrm{true}}$ for that replicate.

\subsection{Methods compared}

We compared four approaches: the classical win ratio, IPCW win ratio, PWR, and win ratio with multiple imputation. The classical two-layer win ratio was computed deterministically by comparing death first and hospitalization second. If either subject had missing hospitalization data, the pair was treated as tied at the hospitalization layer.

For the IPCW win ratio method, hospitalization missingness was handled by a complete-case restriction, and censoring weights were computed within each treatment arm using a Kaplan-Meier estimator of the censoring survival function $G_a(t)=\Prb(C\geq t\mid A=a)$ based on $C_i$. Pairwise contributions were weighted by $1/\{G_1(t)G_0(t)\}$ evaluated at the decisive comparison time.

For the proposed PWR, we estimated layer-specific marginal survival models for death and hospitalization and converted the fitted marginal survival curve for each subject into a conditional survival curve consistent with the observed coarsening state at that layer. Pairwise win, loss, and tie probabilities were computed by numerical integration on $[0,\tau]$, and the hospitalization-layer contribution was propagated by the death-layer tie probability. We considered both a parametric Weibull working model and a Cox working model. All working models included covariates and risk class: $A,X_1,X_2$, and $G$.

For the win ratio with multiple-imputation approach, we imputed hospitalization times only. A parametric survival model was fit for hospitalization including $A,X_1,X_2,G$. Missing hospitalization times were sampled from the fitted model, censored hospitalizations were sampled from the conditional tail beyond the censoring time, and the imputed hospitalization time was truncated at $\min(\widetilde T^{(1)},\tau)$ to respect hierarchy. The classical two-layer win ratio was then computed on each imputed dataset and combined using the geometric mean of the imputed win ratios.

\subsection{PWR working models}

The estimator $\widehat{\theta}_{\PWR}$ requires a working model for the layer-specific marginal survival distributions used to compute conditional win, loss, and tie probabilities under coarsening. In the two-layer setting, we fit separate marginal survival models for death and hospitalization. For death, we fit a Weibull accelerated failure time model using \texttt{survreg} with covariates $(A_i,X_i,G_i)$. For hospitalization, we fit the same working model on the subset of subjects with observed hospitalization data, treating hospitalization as a marginal time-to-event endpoint under the hierarchical censoring rule. As a robustness check, we also considered a Cox proportional hazards working model for each layer, again including $(A_i,X_i,G_i)$ as regressors.

Let $\widehat S_{i1}(t)$ and $\widehat S_{i2}(t)$ denote the fitted marginal survival curves for subject $i$ at layers 1 and 2. To reflect the observed coarsening state, each $\widehat S_{ik}$ is converted to a conditional curve $\widehat S_{ik}(t\mid \mathcal{F}_{ik})$. If an event is observed at time $t_{ik}$, then $\widehat S_{ik}(t\mid \mathcal{F}_{ik})=I(t<t_{ik})$. If the endpoint is right-censored at $c_{ik}<\tau$, then
\[
\widehat S_{ik}(t\mid \mathcal{F}_{ik})
=I(t\leq c_{ik})+I(t>c_{ik})\frac{\widehat S_{ik}(t)}{\widehat S_{ik}(c_{ik})}.
\]
If the subject is known event-free through $\tau$, then $\widehat S_{ik}(t\mid \mathcal{F}_{ik})=1$ for $t\leq \tau$. For missing hospitalization, no conditioning is applied and the fitted marginal curve is used.

\subsection{Simulation setup and performance metrics}

We considered sample size $n=150$ per replicate and fixed the analysis horizon at $\tau=5$. We varied the missingness rate on hospitalization $p_M\in\{0,0.1,0.2,0.3,0.4,0.5\}$ and the dropout rate $p_D\in\{0,0.1,0.2,0.3,0.4,0.5\}$, yielding 36 scenarios. For each scenario, we generated 5000 Monte Carlo replicates and computed each estimator. For each method and scenario, we report percent bias and mean squared error.

\subsection{Results}

Across censoring scenarios, the simulation results reveal systematic differences in performance among the competing win-ratio estimators as missingness in the lower-priority layer increases.

When missingness is absent or mild, all methods perform comparably, exhibiting negligible bias and similar MSE. As the proportion of missing hospitalization data increases, however, the classical win ratio displays progressively larger negative bias, reflecting the increasing frequency with which partially observed pairs are treated as ties. This bias grows monotonically with missingness across all levels of dropout and is accompanied by substantial inflation in MSE.

The IPCW estimator reduces bias relative to the classical approach but remains sensitive to increasing missingness, particularly under moderate to high censoring. In these settings, IPCW exhibits noticeable negative bias and increasing variance, consistent with the loss of efficiency induced by complete-case restriction and inverse-probability weighting when the effective sample size is reduced.

The MI approach shows a contrasting pattern: bias is positive and increases with missingness, indicating systematic overestimation of the win ratio when hospitalization times are imputed from a working survival model. Although MI can yield lower variance than IPCW in some scenarios, this variance reduction does not offset the accumulating bias, resulting in inferior MSE performance at higher levels of missingness.

In contrast, the PWR methods remain stable across the full range of missingness and censoring conditions. Both the parametric and Cox-based implementations exhibit near-zero bias throughout, with only modest increases in MSE as missingness increases. Once missingness exceeds approximately 20\%, PWR consistently achieves the smallest MSE among all methods, and its performance is largely insensitive to the degree of dropout. The close agreement between the parametric and Cox versions further suggests robustness to working-model specification.

\begin{figure}[htbp]
\centering
\includegraphics[width=\linewidth]{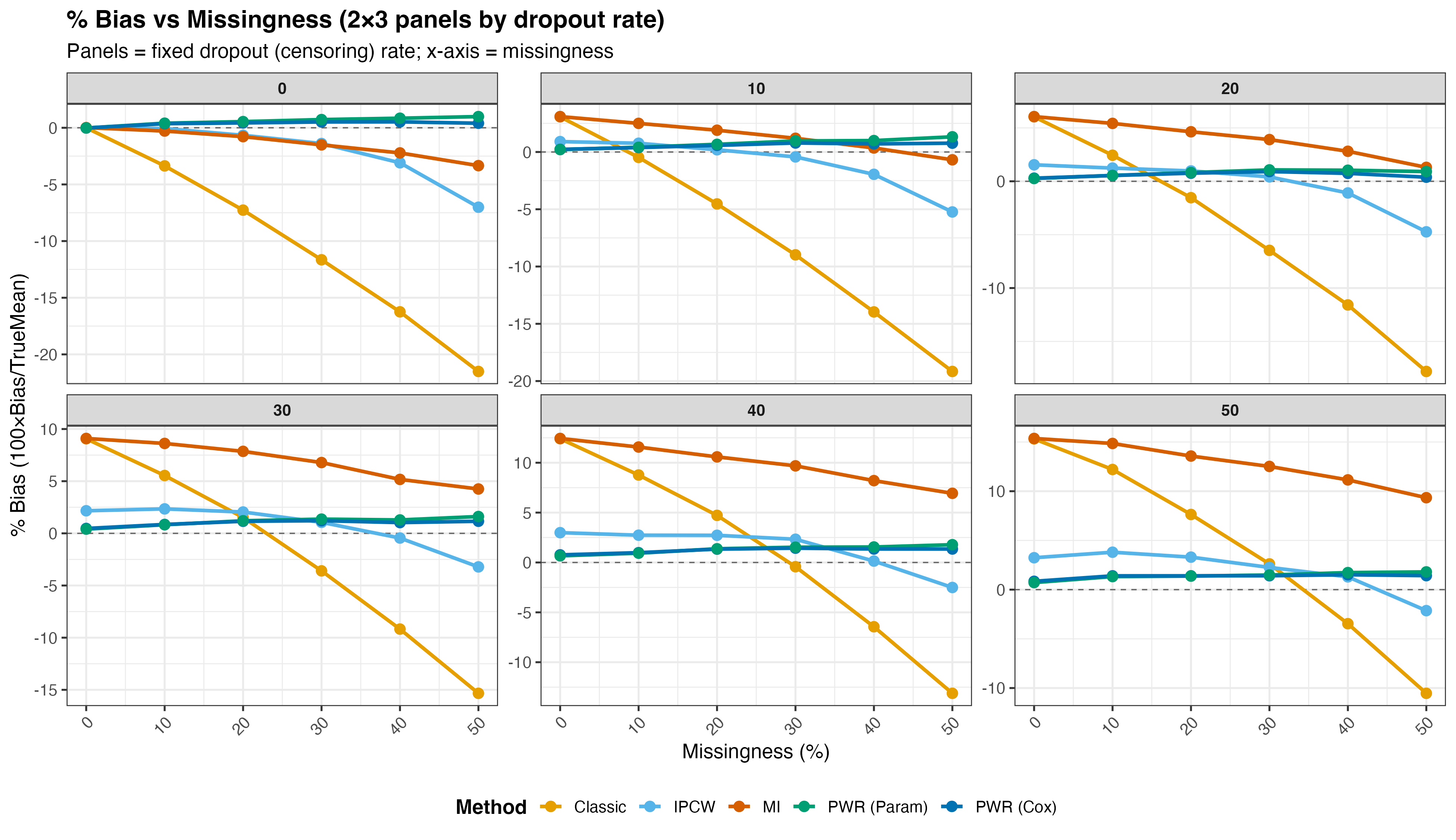}
\caption{Percent bias comparison across missingness and dropout settings.}
\label{fig:bias}
\end{figure}

\begin{figure}[htbp]
\centering
\includegraphics[width=\linewidth]{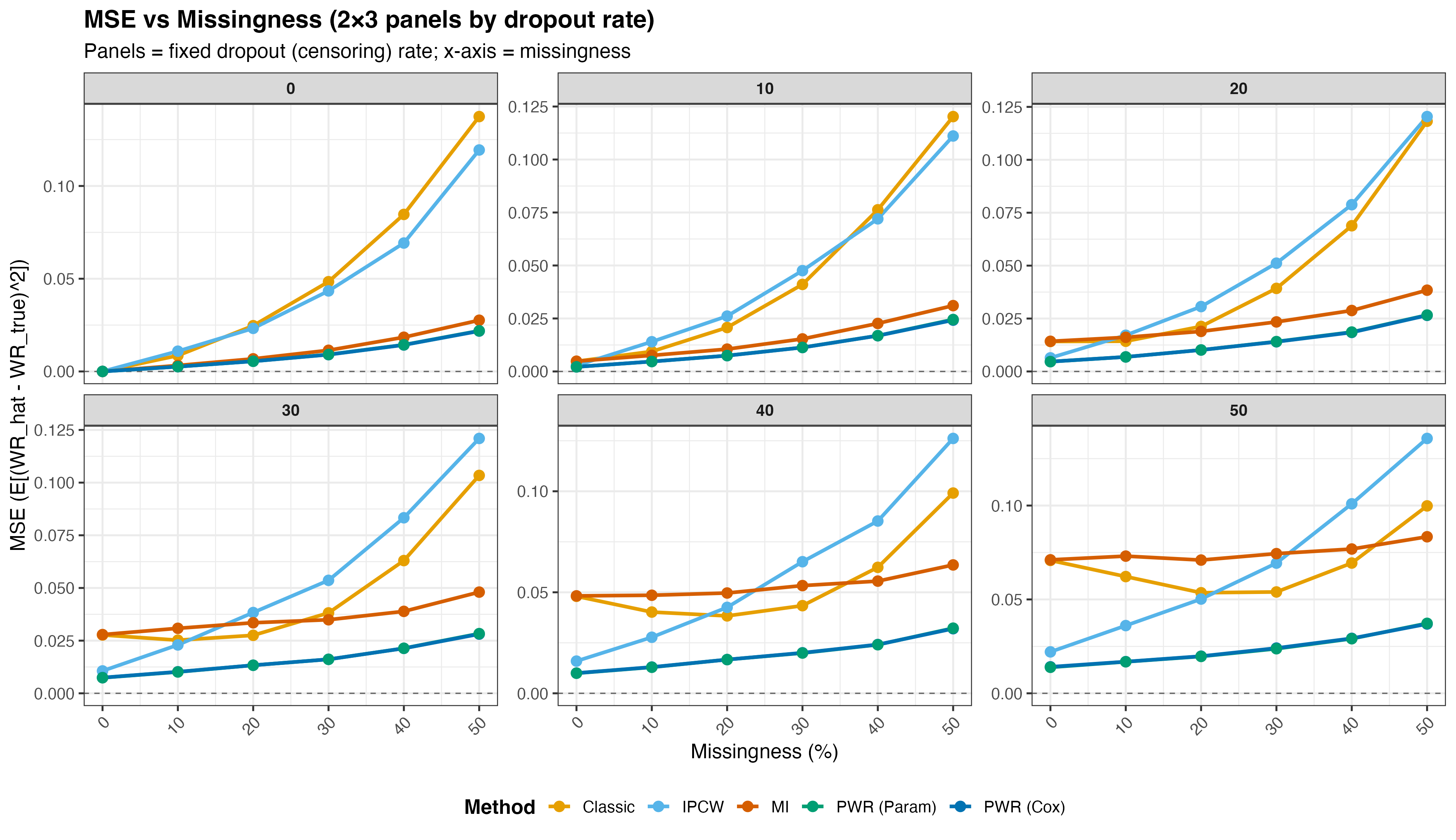}
\caption{Mean squared error comparison across missingness and dropout settings.}
\label{fig:mse}
\end{figure}

\begin{figure}[htbp]
\centering
\includegraphics[width=\linewidth]{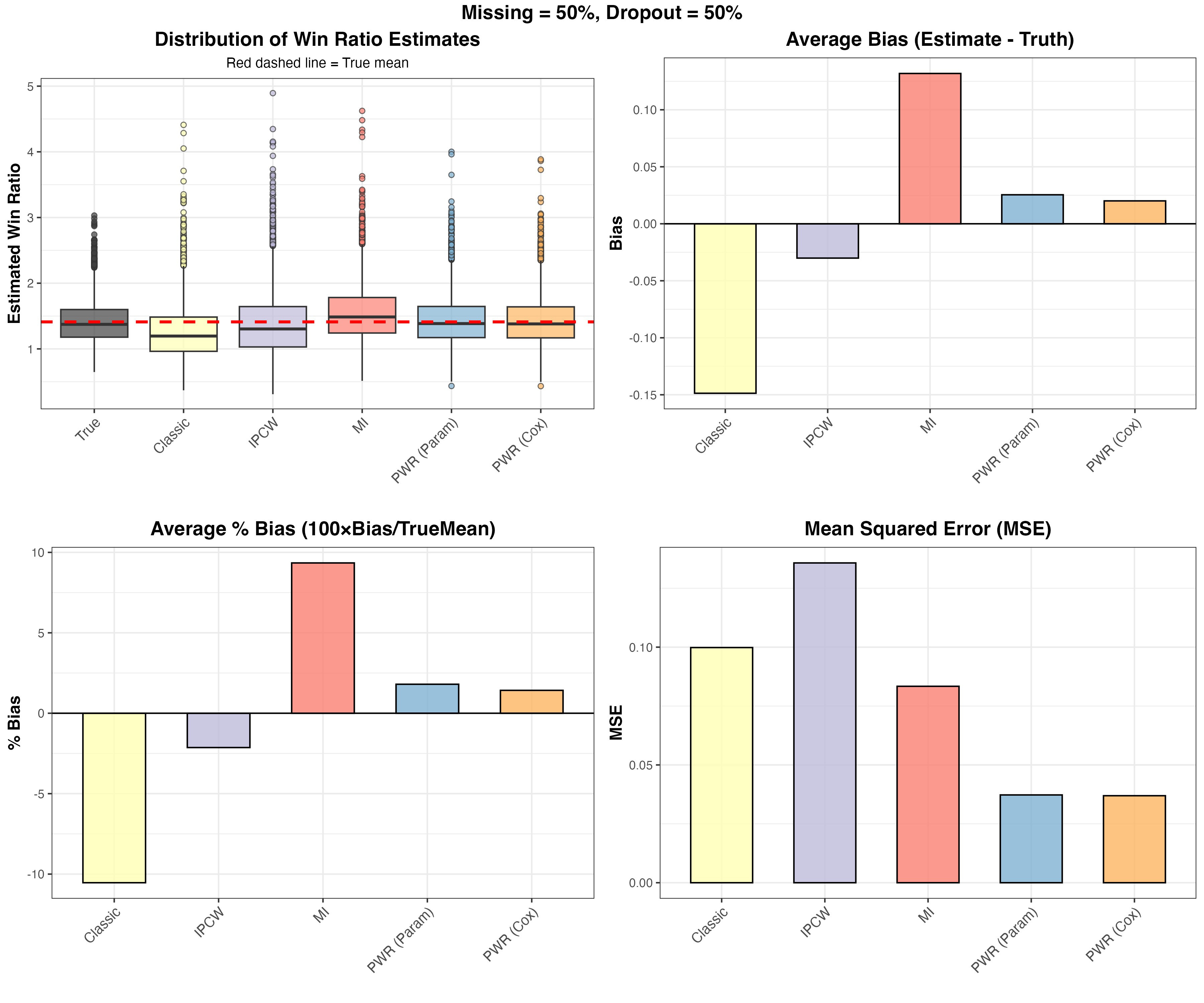}
\caption{Comparison across win ratio methods under 50\% missingness and 50\% dropout.}
\label{fig:high_missingness}
\end{figure}

\section{Case studies}
\label{sec:cases}

We illustrate the proposed PWR using two clinical trial datasets chosen to represent complementary data regimes commonly encountered in practice. The first example, based on the primary biliary cirrhosis (PBC) trial, features minimal censoring prior to the analysis horizon, providing a calibration setting in which standard and probabilistic win-ratio methods are expected to agree closely. The second example, based on the HF-ACTION trial, exhibits heavy right censoring prior to the analysis horizon while retaining minimal endpoint-specific missingness.

\subsection{Primary biliary cirrhosis study}

We illustrate the PWR using data from the PBC trial available in the R \texttt{survival} package. The study randomized patients to D-penicillamine or placebo and followed them for mortality, liver transplantation, or censoring. The original trial found no clear survival benefit of treatment.

We constructed a two-layer prioritized composite endpoint. Layer 1 was time to death, and layer 2 was time to liver transplantation. Death was treated as the highest-priority event; transplantation was considered only for pairs unresolved at the death layer. An administrative analysis horizon $\tau=2000$ days was imposed uniformly across all methods.

We compared four approaches: the classical two-layer win ratio, PWR using Cox models, PWR using Weibull models, and IPCW win ratio. For PWR, marginal survival models were fit separately for death and transplantation, including treatment and baseline covariates (age, sex, bilirubin, albumin, and edema). Fitted marginal survival curves were converted into conditional survival curves consistent with each subject's observed coarsening state, and pairwise win/loss probabilities were obtained by numerical integration. Layer 2 contributions were propagated by the tie probability from layer 1.

Point estimates of the win ratio were accompanied by nonparametric bootstrap standard errors and confidence intervals computed on the log win ratio scale. Bootstrap resampling was performed at the patient level with $B=300$ replicates.

\begin{table}[htbp]
\centering
\caption{Win ratio estimates for the PBC dataset using a two-layer hierarchical endpoint (death, then liver transplant). Standard errors and confidence intervals are computed on the log scale via the nonparametric bootstrap.}
\label{tab:pbc}
\begin{tabular}{lccc}
\toprule
Method & Estimate & SE\{$\log(\WR)$\} & 95\% CI (WR)\\
\midrule
Classic WR & 1.04 & 0.21 & (0.68, 1.55)\\
PWR (Cox) & 1.06 & 0.19 & (0.76, 1.53)\\
PWR (Weibull) & 1.06 & 0.20 & (0.71, 1.55)\\
IPCW WR & 1.00 & 0.21 & (0.69, 1.61)\\
\bottomrule
\end{tabular}
\end{table}

Table~\ref{tab:pbc} shows that across all methods, estimated win ratios are close to 1, with overlapping confidence intervals. No approach suggests a meaningful treatment effect, consistent with the original trial conclusions. Differences between classical WR, IPCW, and PWR are minimal in this dataset, reflecting the absence of endpoint-specific missingness and relatively complete follow-up. This example therefore serves primarily as a calibration exercise, demonstrating that PWR reproduces standard win ratio results under near-complete data while remaining numerically stable under both Cox and parametric working models.

\subsection{HF-ACTION case study}

We illustrate the proposed PWR using individual patient data from the HF-ACTION trial, a randomized study of exercise training in patients with chronic heart failure. The trial was originally analyzed using a Cox proportional hazards model for time to first occurrence of all-cause mortality or all-cause hospitalization. In the unadjusted analysis, no statistically significant treatment effect was observed for the primary composite endpoint, while a modest benefit emerged only after adjustment for prespecified prognostic covariates. No significant reduction in all-cause mortality alone was reported. Here, HF-ACTION is used as a methodological calibration example rather than a re-analysis of the original clinical estimand.

We constructed a two-layer prioritized composite endpoint, with time to death as the highest-priority layer and time to first hospitalization as the second layer. The hospitalization layer was evaluated only for subject pairs unresolved at the death layer, enforcing the hierarchical stopping rule. Recurrent hospitalizations beyond the first event were not considered.

Rather than fixing a common calendar follow-up time, inference was conducted using a $\tau$-restricted estimand, where $\tau$ was defined as the 95\% quantile of the subject-level follow-up distribution. Subjects observed event-free through $\tau$ are fully observed with respect to the estimand, even though their follow-up may extend beyond $\tau$. In contrast, censoring or hierarchical truncation occurring prior to $\tau$ induces coarsening and contributes fractional win, loss, and tie probabilities in the PWR.

In HF-ACTION, censoring prior to $\tau$ was substantial: 310 of 426 subjects (72.8\%) were censored before $\tau$ at the death layer, and 105 of 426 (24.6\%) were censored before $\tau$ at the first-hospitalization layer. Uncertainty penalties in the PWR therefore arise from censoring before $\tau$, while administrative censoring beyond $\tau$ does not affect the $\tau$-restricted comparison.

We compared the classical two-layer win ratio, PWR using Cox working models, PWR using Weibull working models, and IPCW-adjusted win ratio. For the PWR, marginal survival models were fit separately for death and first hospitalization. Fitted marginal survival curves were converted into conditional survival functions consistent with each subject's observed coarsening state, and pairwise win, loss, and tie probabilities were computed by numerical integration over $[0,\tau]$. Contributions from the hospitalization layer were propagated by the estimated tie probability from the death layer. Inference was performed on the log win ratio scale using nonparametric bootstrap resampling at the subject level.

\begin{table}[htbp]
\centering
\caption{Win ratio estimates for the HF-ACTION dataset using a two-layer hierarchical endpoint (death, then first hospitalization). Standard errors and confidence intervals are computed on the log scale via the nonparametric bootstrap.}
\label{tab:hfaction}
\begin{tabular}{lccc}
\toprule
Method & Estimate & SE\{$\log(\WR)$\} & 95\% CI (WR)\\
\midrule
Classic WR & 1.26 & 0.12 & (1.00, 1.60)\\
PWR (Cox) & 1.30 & 0.13 & (1.02, 1.72)\\
PWR (Weibull) & 1.30 & 0.13 & (1.03, 1.69)\\
IPCW WR & 1.25 & 0.12 & (0.99, 1.56)\\
\bottomrule
\end{tabular}
\end{table}

Table~\ref{tab:hfaction} shows that estimated win ratios were similar across methods, ranging from 1.25 to 1.30, with substantial overlap in confidence intervals. The PWR implemented using Cox and Weibull working models yielded nearly identical point estimates, indicating stability with respect to the choice of working model. Relative to the classical and IPCW-adjusted win ratios, the PWR produced slightly larger point estimates, resulting in confidence intervals that were shifted upward. These differences were modest, reflecting the fact that mortality and hospitalization outcomes in HF-ACTION are largely complete and subject primarily to right censoring rather than endpoint-specific missingness.

Importantly, the PWR did not induce spurious effects or inflated precision. Its agreement with existing win-ratio methods in this setting demonstrates that the proposed framework generalizes the classical win ratio without altering its behavior when data are nearly complete, while remaining well defined under censoring.

\section{Concluding remarks}
\label{sec:discussion}

The win ratio provides an appealing framework for analyzing prioritized composite endpoints, but standard implementations rely on deterministic pairwise comparisons and are not designed to accommodate censoring and layer-specific missingness. As a result, classical win ratio analyses may lose efficiency or exhibit bias when comparisons are only partially observed, particularly for lower-priority outcomes.

We propose the PWR, which recasts hierarchical comparisons in terms of conditional win, loss, and tie probabilities given the observed coarsening state. By modeling marginal outcome distributions and propagating uncertainty across layers solely through tie probabilities, the PWR preserves the clinical priority structure of the win ratio while providing a coherent treatment of censoring and missingness. Partially observed comparisons are automatically downweighted through these conditional probabilities, yielding a data-adaptive penalization of comparisons with limited information rather than treating them as full ties or discarding them entirely. The resulting estimator is well defined and reduces exactly to the classical win ratio estimator when outcomes are fully observed.

Simulation studies demonstrated that the PWR remains stable across a wide range of missingness and censoring scenarios, with reduced bias and MSE relative to classical, IPCW-based, and multiple-imputation approaches when lower-priority outcomes are incompletely observed. Parametric and Cox-based implementations of the PWR performed similarly, suggesting robustness to working-model choice. Analysis of the PBC trial illustrated that, in settings with minimal missingness, the PWR yields results comparable to existing win ratio methods and aligns with established trial conclusions.

Overall, the PWR offers a principled and flexible extension of the win ratio for clinical trials with prioritized composite endpoints. By separating the clinical comparison rule from uncertainty induced by coarsened observation, the PWR provides a robust alternative for inference in the presence of censoring and missingness.

\appendix
\section{Regularity conditions}
\label{app:regularity}

This appendix states assumptions required for identification, consistency, and stable large-sample behavior of the PWR estimator for hierarchical time-to-event endpoints with censoring and missingness. We consider a general hierarchy of $K\geq2$ ordered endpoints. For each subject, let
\[
\{T^{(1)},T^{(2)},\ldots,T^{(K)}\}
\]
denote the potential event times corresponding to $K$ prioritized endpoints, where lower indices indicate higher priority. Let $C$ denote a censoring time, $A\in\{0,1\}$ the treatment indicator, $X$ baseline covariates, and $M^{(k)}\in\{0,1\}$ missingness indicators for the $k$th endpoint. All estimands are defined on a fixed analysis horizon $[0,\tau]$.

\subsection*{A. Structural and causal assumptions}

\paragraph{A1. Randomization or ignorability.}
Treatment assignment is randomized and independent of baseline covariates and latent potential event-time processes at baseline; subjects are independent within arms.

\paragraph{A2. Consistency.}
The observed event time for each endpoint equals the potential event time under the treatment actually received. For each endpoint $k$,
\[
T^{(k)}=T^{(k)}(A),
\]
where $T^{(k)}(a)$ denotes the potential event time under treatment $a$.

\paragraph{A3. Hierarchical compatibility.}
Lower-layer observed time is truncated or censored at the first precluding higher-layer event.

\subsection*{B. Censoring assumptions}

\paragraph{B1. Independent censoring.}
Conditional on treatment and baseline covariates, censoring is independent of all potential event times:
\[
C\perp\!\!\!\perp \{T^{(1)}(a),\ldots,T^{(K)}(a)\}\mid (A=a,X),
\qquad a\in\{0,1\}.
\]

\paragraph{B2. Positivity of censoring survival.}
There exists $\epsilon>0$ such that
\[
\Prb(C\geq t\mid A=a,X)\geq \epsilon
\]
for all $t\in[0,\tau]$ and $a\in\{0,1\}$.

\paragraph{B3. Estimability of the censoring distribution.}
The censoring survival functions $G_a(t)=\Prb(C\geq t\mid A=a)$ are consistently estimable, for example via Kaplan-Meier or Cox regression. Mild model misspecification affects efficiency but not first-order consistency under Assumption B1.

\subsection*{C. Missingness assumptions}

\paragraph{C1. Layer-specific missingness.}
Missingness may occur in any subset of layers $\mathcal{K}_M\subseteq\{1,\ldots,K\}$. For a given layer $k\in\mathcal{K}_M$, missingness of $(T^{(k)},\Delta^{(k)})$ is allowed only among subjects whose outcomes at all higher-priority layers $\ell<k$ are unresolved at the time layer $k$ would be observed.

\paragraph{C2. Missing at random within layers.}
For each layer $k\in\mathcal{K}_M$,
\[
M^{(k)}\perp\!\!\!\perp T^{(k)}
\mid
(A,X,T^{(1)}\geq t^{(1)},\ldots,T^{(k-1)}\geq t^{(k-1)}),
\]
where $t^{(j)}$ denotes the observed follow-up time in layer $j$.

\paragraph{C3. No informative missingness in higher-priority layers.}
For layers that determine eligibility for lower-priority comparisons, missingness does not induce informative selection beyond what is captured by observed covariates and prior-layer survival.

\subsection*{D. Modeling assumptions for PWR}

\paragraph{D1. Marginal survival modeling.}
For each layer $k$, a working marginal survival model
\[
S_k(t\mid A,X)=\Prb(T^{(k)}>t\mid A,X)
\]
is specified and estimable. No joint modeling across layers is performed.

\paragraph{D2. Correct specification of the first non-tied layer.}
For consistency of the PWR estimator, the marginal survival model must be correctly specified for all layers with nonzero asymptotic reach probability. The highest-priority layer is always active and must be correctly specified. Misspecification in lower-priority layers affects efficiency and may induce bias unless those layers are reached with asymptotically negligible probability.

\paragraph{D3. Marginal modeling of lower-priority layers.}
Lower-priority layers are modeled marginally, without conditioning on higher-priority event times. Dependence across layers is incorporated solely through explicit multiplication by the estimated tie probabilities from higher-priority layers.

\subsection*{E. Pairwise comparison and weighting assumptions}

\paragraph{E1. Sequential tie propagation.}
Pairwise comparisons proceed hierarchically. Contributions from layer $k$ enter the PWR estimand only when all higher-priority layers $1,\ldots,k-1$ result in ties.

\paragraph{E2. Finite pairwise moments.}
The weighted pairwise win and loss indicators have finite second moments, permitting application of U-statistic theory.

\subsection*{F. Regularity conditions}

\paragraph{F1. Independent subjects.}
Subjects are independent and identically distributed.

\paragraph{F2. Bounded time horizon.}
All event times are evaluated on a compact interval $[0,\tau]$ with $\tau<\infty$.

\paragraph{F3. Regular survival functions.}
For each layer $k$, the survival function $S_k(t\mid A,X)$ is right-continuous, non-increasing, and bounded away from zero on $[0,\tau]$.

\section{Asymptotic properties of the PWR estimator}

Let $\widehat{\theta}_{\PWR,n}$ denote the PWR estimator based on $n$ independent subjects.

\paragraph{Consistency.}
Under Assumptions A--F, the PWR estimator is consistent:
\[
\widehat{\theta}_{\PWR,n}\stackrel{p}{\longrightarrow}\theta.
\]
Consistency follows from uniform consistency of the estimated marginal survival functions for each layer, stability of the conditional coarsening operators defined by observed event, censoring, and missingness states, and the law of large numbers for pairwise U-statistics. Correct specification of the marginal model for the highest-priority layer that determines pairwise ordering ensures identification of the leading contribution to $\theta$, while lower-priority layers enter only through tie-propagated terms.

\paragraph{Asymptotic normality.}
Under the same assumptions and mild smoothness conditions on the working survival models, the estimator admits U-statistic-type asymptotic normality:
\[
\sqrt{n}(\widehat{\theta}_{\PWR,n}-\theta)
\stackrel{d}{\longrightarrow}
N(0,\sigma^2),
\]
for a finite variance $\sigma^2$ that depends on the joint distribution of the pairwise comparison indicators and the estimation error of the marginal survival functions. The asymptotic variance can be consistently estimated using resampling or sandwich-type methods.

\paragraph{Proof sketch.}
The estimator $\widehat{\theta}_{\PWR,n}$ can be written as a smooth functional of empirical averages of pairwise win, loss, and tie indicators with estimated nuisance parameters. After conditioning on the observed coarsening states, the estimator reduces to a ratio of U-statistics with plug-in estimates of layer-specific survival functions. Standard results for U-statistics with estimated nuisance parameters imply asymptotic linearity, provided that the marginal survival estimators converge uniformly and that censoring and missingness mechanisms satisfy Assumptions B and C.

\end{document}